\documentclass[aps,prb,twocolumn]{revtex4}
\usepackage{graphicx}

\begin{document}
\title{Correlation effects on the electronic structure  of TiOCl: a NMTO+DMFT study}

\author{T. Saha-Dasgupta$^1$, A. Lichtenstein$^{2}$ and  R. Valent\'\i$^{3}$} 

\address{$^1$S.N. Bose National Centre for Basic Sciences, 
JD Block, Sector III,
 Salt Lake City, Kolkata 700098, India.}

\address{$^2$ Institut f\"ur Theoretische Physik,
Universit\"at Hamburg,20355 Hamburg, Germany.}

\address{$^3$ Institut f\"ur Theoretische Physik,
J.W.Goethe-Universit\"at Frankfurt,
60054 Frankfurt/Main, Germany.}

\date{\today}

\begin{abstract}

%We present an  \textit{ab initio} electronic structure calculation
%of the correlated, quantum spin compound, TiOCl in the high temperature phase. 
Using the recently developed N-th order muffin-tin orbital-based downfolding 
technique in combination with the Dynamical Mean Field theory, we
investigate the electronic properties of the much discussed Mott insulator TiOCl
in the undimerized phase. Inclusion of correlation effects through this
approach provides a description of the spectral function into an upper
 and a lower Hubbard band with broad valence states 
%the insulating state of TiOCl was investigated and compared with
%experimental results.
formed out of the orbitally 
polarized, lower Hubbard band.  We find that these results are in good agreement with  recent 
photo-emission spectra.

\end{abstract}
\pacs{71.27.+a, 71.30.+h, 71.15.Ap} 
%%\pacs{75.30.Gw, 75.10.Jm, 78.30.-j}

\maketitle

%%%%%%%%%%%%%%%%%%%%%%%%%%%%%%%%%%%%%%%%%%%%%%%%%%%%%%%%%%%%%%%%%%%
%%%%%%%%%%%%%%%%%%%%%%%%%%%%%%%%%%%%%%%%%%%%%%%%%%%%%%%%%%%%%%%%%%%

%\vspace*{1cm}

{\it Introduction -}

A challenging task in the field of strongly electron-correlated
materials
is the description of correlation effects, within a possibly controlled
approximation, since these effects are essential in order to understand
the behavior of these materials.  In the past years there has been an
enormous
effort into improving this description by combining {\it ab-initio} 
calculations with many-body methods like the LDA+U (Local Density
Approximation
plus on-site U)\cite{Anisimov_91} and the LDA+DMFT (Local Density
Approximation
combined with Dynamical Mean Field Theory)\cite{review_dmft}.
While the original implementation of the Dynamical Mean Field theory
was based on a  single-band model, most compounds of interest involve more
than one correlated orbital. This calls for the need of a multi-orbital
extension of the LDA+DMFT technique.
 Recently,  a new implementation of the multi-orbital LDA+DMFT has been
 proposed  which uses   
 the localized Wannier functions generated by the N-th order muffin-tin-orbital
 (NMTO) method\cite{nmto}
 in order to construct the LDA Hamiltonian and solves the many-body problem by DMFT 
including the non-diagonal contribution of the on-site self-energy. Such approach has been
found to be highly successful in a series of recent applications 
\cite{Pavarini_04, Poteryaev_04, Biermann}. We will consider this
procedure here
 in order to unveil the electronic properties of the  layered Mott
 insulator TiOCl. 
%and discuss its improvements with respect to other
% LDA+DMFT
%implementations.

\begin{figure}
\includegraphics[width=7cm,keepaspectratio]{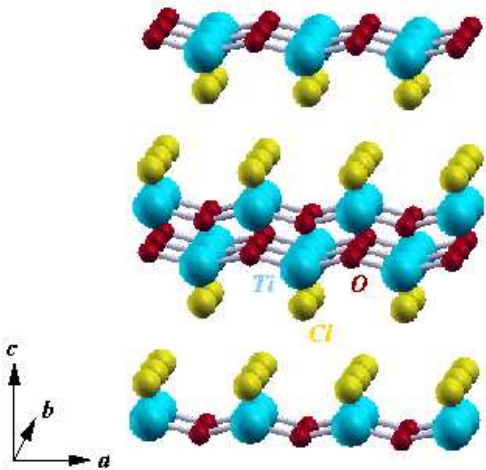}
\caption{Layered structure of TiOCl crystal. Two O atoms from the same layer
and two O and two Cl atoms from neighboring layers form the O$_{4}$Cl$_{2}$ 
octahedra surrounding the Ti atom. }
\label{structure}
\end{figure}

 % The layered Mott insulator 
TiOCl,  which consists of  bilayers of Ti$^{3+}$
and O$^{2-}$ parallel to the $ab$ plane, separated by layers of Cl$^{-}$
ions stacked along the $c$-axis (see Fig.\ \ref{structure}) has recently raised a lot of
  discussion
 due to its puzzling behavior at moderate to low temperatures. 
Large phonon anomalies have been observed in Raman measurements\cite{lemmens_03_2}
 at temperatures around 135K as well as
 temperature-dependent
g-factors and line-widths in ESR\cite{Kataev_03}. The susceptibility
 measurements\cite{Seidel_03} show  a kink at T$_{c2}$ = 94 K and an exponential drop
at $T_{c1}$ = 66 K, indicating the opening of a spin-gap which is
accompanied by a doubling of unit-cell along the $b$-axis\cite{Shaz_04}.
While  phonons are undoubtedly playing an important role in the behavior
of
 this system, electron correlation is  fundamental in order to understand
these anomalous properties and  we will here concentrate on the electronic description.

 Recent  electronic structure calculations of TiOCl within the framework of 
 LDA+U \cite{Seidel_03, Saha_04} for the crystallographic data at T $>$ T$_{c2}$ 
 show  that in this temperature range
the ground state of the system is described by  chains of Ti ions
along the $b$-direction (see Fig.\ \ref{structure}) with the 3$d$ 
electrons occupying the $d_{xy}$ 
orbitals. Such a study \cite{Saha_04} further revealed the importance of the phonon degrees
of freedom at temperatures  T $>$ T$_{c2}$ by pointing out
that certain A$_{g}$ phonon modes consistent with the $Pmmn$ space group
of the high temperature structure may lead to orbital fluctuations
%at high temperatures 
with the ground state switching from $d_{xy}$ to $d_{yz}/d_{xz}$. However, the 
implementation of LDA+U calculations \cite{review_ani} involves the  assumption of a particular 
spin ordering in the system, which is fictitious since there is no true long range 
order in  TiOCl. This makes the comparison of the LDA+U 
derived DOS with the experimentally measured photo-emission data doubtful.
Also, LDA+U being a mean-field method,  treats the correlation effects beyond LDA
in a mean-field sense and suppresses all the fluctuation
% {\bf contributions}
 effects. For early transition metal oxides like titanates which are
moderately correlated systems such fluctuation effects can turn out to be essentially
important. One  ingenious way of treating the dynamical fluctuation effects, 
 is the  DMFT
which - though it freezes the spatial fluctuations-   takes fully into account
the temporal fluctuations. In the present communication, we aim
to study the electronic structure of the high temperature ( T $>$ T$_{c2}$) 
phase of TiOCl by means of the NMTO+DMFT method.

%===========================================================================
%===========================================================================

%\vskip .1in

{\it Methodology -} 

The octahedral crystal field provided by the O$_{4}$Cl$_{2}$ octahedron, 
surrounding the Ti site splits the five degenerate d-orbitals into 
$t_{2g}$ and $e_g$ manifolds. The Ti$^{3+}$ ion, in a 3d$^{1}$ configuration,
therefore fills 1/6-th of the $t_{2g}$ complex, leaving the $e_g$ complex
empty. The basic LDA electronic structure was reported in Ref.\cite{Seidel_03,Saha_04}. The 
essential features involve Cl and O$-p$ dominated bands extending from about -8 to -4 eV,
separated by a gap of about 2.5 eV from the Ti-$d$ complex, with 
the zero of energy
set at the Fermi level. The Ti-$t_{2g}$ bands cross the Fermi level with a tiny gap
of about 0.2 eV between the occupied $t_{2g}$ manifold and unoccupied 
$e_g$ manifold. 

The TiOCl system can therefore be described by a low-energy, multi-band
Hubbard Hamiltonian, 

\begin{eqnarray}
H & = & H^{LDA} + \frac{1}{2} \sum_{imm'\sigma} U_{mm'} n_{im\sigma} n_{im'-\sigma} \\ \nonumber
& + & \frac{1}{2} \sum_{im(\ne m')\sigma} (U_{mm'} - J_{mm'}) n_{im\sigma} n_{im'\sigma} 
\label{hamiltonian}
\end{eqnarray}

where $ n_{im\sigma} = c_{im\sigma}^{\dagger}c_{im\sigma}$ and  $c_{im\sigma}^{\dagger}$
creates a $\sigma$-spin electron in a localized $t_{2g}$ orbital ($m$) at site $i$. 
$H^{LDA}$ is the one-electron %{\bf contribution} 
 Hamiltonian  given by LDA. We assume double counting corrections to
be orbital-independent within the $t_{2g}$ manifold, thus resulting into a simple shift
of the chemical potential. The many-body Hamiltonian defined in 
Eq. 1
 depends on the
choice of the localized orbitals. In the present calculation, we employed the NMTO \cite{nmto}
method for defining these localized orbitals. In this method, a basis set of localized orbitals 
is cons\-truc\-ted from the exact scattering solutions 
 of a superposition of short-ranged, spherically-symmetric potential
wells (the so-called muffin-tin approximation to the potential)
at a mesh of energies, $\epsilon_{0}, \epsilon_{1}, 
\ldots, \epsilon_{N}$. The number of energy points, N, 
defines the order of such muffin-tin orbitals, the NMTO's. Each NMTO satisfies a 
specific boundary condition which provides it with an  orbital character and makes it 
localized. The NMTO's being energy-selective in nature are flexible and may be chosen to
span selected bands. If these bands are isolated -as is in the case of $t_{2g}$-derived bands
in TiOCl-  the NMTO set spans the Hilbert space of the Wannier functions. In other words, the
orthonormalized NMTO's are the localized Wannier functions. In our example of TiOCl, the low-energy
Hamiltonian defined in Eq. 1, involves only three correlated, localized $t_{2g}$ Wannier orbitals and
no other orbitals.  However, for such Wannier functions to be
 complete, they  must
involve contributions from the other degrees of freedom, {\it e.g.} O${-p}$ and Cl${-p}$. This is
achieved via the {\it downfolding} procedure \cite{nmto} which adds on to Ti centered $t_{2g}$ orbitals,
  tails with O${-p}$ and Cl${-p}$ character, thereby defining an effective set of $t_{2g}$
orbitals. It is important to note here, that  the choice of the orbital
symmetries $d_{xy}, d_{yz}, \ldots, d_{3z2-1}$ depends
on the choice of the co-ordinate system. Throughout this paper and also in previous 
 work\cite{Seidel_03, Saha_04},
the choice has been made as $\hat{z}$ = $a$, and $\hat{x}$ and $\hat{y}$ axes
rotated 45$^{o}$ with respect to $b$ and $c$. However, due to the distorted geometry of
the TiO$_{4}$Cl$_{2}$ octahedra, the chosen co-ordinate system differs from the local co-ordinate
system and the various $d$ orbitals  defined following the above mentioned
co-ordinate system are not the exact eigenstates of the on-site Ti$-d$ Hamiltonian. This introduces mixing
between the $t_{2g}$ and $e_g$ symmetries. The $t_{2g}$ NMTO's therefore, contain
also the on-site and tail $e_g$ character. For a realistic
description of the LDA contribution and for the proper description of
the localized orbitals in Eq. 1 
it is essential to consider these issues. To our knowledge, so far no other 
LDA+DMFT is capable of taking these contributions into account properly. The plot of 
such Wannier functions is shown in Ref.\ \cite{Saha_04}.

\begin{figure}

\includegraphics[width=5cm,keepaspectratio,angle=0]{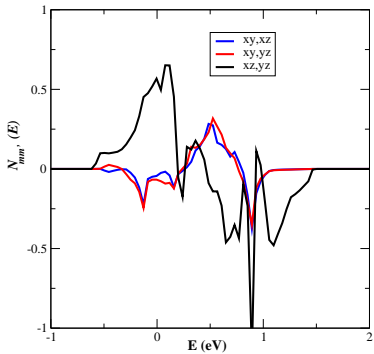}

\vspace*{0.4cm}
\caption{ Off-diagonal matrix elements of $t_{2g}$ LDA DOS matrix (states/eV) 
in the Wannier representation.}
\label{partial}
\end{figure}

In Fig.\ \ref{partial}, we show the off-diagonal part of the $3\times3$ density matrix constructed
out of the effective, downfolded $t_{2g}$ NMTO's:

\[
N_{mm'} = \sum_{k,n} u_{k,n}^{m} \delta (E - E_n,k) u_{k,n}^{m'}
\]

$m, m'$=1,2,3 refer to $t_{2g}$ orbitals at the same site and
$u_{k,n}^{m}$ are the
appropriate normalized eigenvectors for the downfolded $t_{2g}$ bands in the 
orthonormalized NMTO basis. The summation over $k$ runs over all the $k-$points in the Brillouin
zone (BZ) and $n$ runs over all the $t_{2g}$ bands. As is seen from the plot, it has non-negligible
off-diagonal elements within the chosen co-ordinate system.

The DMFT \cite{review_dmft} maps the many-body crystal problem defined in Eq. 1 onto an effective, 
self-consistent, multi-orbital quantum impurity problem. The corresponding local Green's function matrix 
is calculated via the BZ integration, 

\begin{equation}
G(\omega_n) = \sum_{k} [ (\omega_n + \mu) I - H^{LDA} (k) - \Sigma (\omega_n) ]^{-1}
\end{equation}

where $\mu$ is the chemical potential defined self-consistently through the total number of
electrons, $\omega_n = (2n+1) \pi/\beta$ are the Matsubara frequencies with $\beta$ as the inverse
temperature ($\beta$ = 1/T). $\Sigma$ is the self-energy matrix.

%===========================================================================
%===========================================================================

In our DMFT implementation we consider all components of the self-energy 
matrix $\Sigma_{mm'}$ between different $t_{2g}$ Wannier functions at a given 
Ti site, including the off-diagonal contribution. The importance of these 
off-diagonal elements has already been inferred from the plot
shown in Fig. 2. This is in contrast  to earlier LDA+DMFT implementations \cite{lda+dmft_old}
where it was
assumed that the on-site block of the single-particle Green's function is diagonal in the space
of the localized orbitals defining the many-body Hamiltonian. 
%%{\bf RV: While such approximations are good for the cubic $t_{2g}$ systems, they loose in accuracy for distorted $t_{2g}$ materials.}
While such an approach is exact for the materials with perfect cubic symmetry, it becomes an approximation
for materials having distortions.
% {\bf with respect to}
%, producing deviation from perfect cubic symmetry.

The self-consistency condition within DMFT implies the local Green's function to be the same as
the corresponding solution of the quantum impurity problem
$G_\sigma (\tau - \tau') = 
1/Z \int D[{\bf c},{\bf c}^{\dagger}] e^{-S_{eff}} {\bf c}(\tau) {\bf c}^{\dagger}(\tau')$, where the effective 
action $S_{eff}$ is defined in terms
of the so-called bath Green's function ${\cal G}^{-1}_{\sigma} (\omega_{n}) = G_{\sigma}^{-1}(\omega_{n}) 
+ \Sigma_{\sigma}(\omega_{n})$, which describes the energy, orbital, spin and temperature dependent 
interaction of a particular site with the rest of the medium,

\begin{eqnarray*}
S_{eff} & = & - \int_{0}^{\beta} d \tau \int_{0}^{\beta} d \tau'  Tr[{\bf c}^{\dagger}(\tau){\cal G}^{-1}_{\sigma}
(\tau, \tau'){\bf c}(\tau')] \\ 
& + & \frac{1}{2} \sum_{imm'\sigma} U_{mm'} n_{im\sigma} n_{im'-\sigma}\\ 
& + & \frac{1}{2} \sum_{im(\ne m')\sigma} (U_{mm'} - J_{mm'}) n_{im\sigma} n_{im'\sigma}
\end{eqnarray*}

$ {\bf c}(\tau) = [c_{im\sigma}(\tau)]$ is the super-vector of the Grassman variables and 
Z is the partition function. The multi-orbital quantum impurity problem is solved by the
numerically exact Quantum Monte Carlo (QMC) scheme. Standard parametrization \cite{param} 
has been used for
the direct and exchange terms of the screened Coulomb interaction, $U_{mm'}$ and $J_{mm'}$
with $U_{mm}$ = U,  $U_{mm'}$ = U - 2J and $J_{mm'(\ne m)}$ = J. For our calculation
we have used U = 4 eV and J = 0.5 eV, which are reasonable choices for an early transition
metal like Ti\cite{Saha_04}. The computational effort becomes prohibitive rather quickly as one lowers the
temperature since in order to  maintain the accuracy of the calculation one needs to increase the imaginary 
time slices as one increases $\beta$. The results reported in the following 
are done for $\beta$ = 20 
(T = 580 K) with 100 slices in imaginary time and 10$^{6}$ QMC sweeps. The maximum entropy
method \cite{max-ent} has been used for analytical continuation of the diagonal part of the Green's function
matrix to the real energy axis to get the DMFT spectral density.

In this context, recently a LDA+DMFT calculation has been
reported for TiOCl \cite{craco}. The calculation scheme in
Ref. \cite{craco} is based on  a multi-orbital generalization
of the iterated perturbation theory (IPT) for impurity solver. We consider the
NMTO+DMFT scheme  
to be superior to the IPT approximation for the present problem. The
multi-orbital QMC solver gives an accurate solution of the correlated
orbitally polarized nonmagnetic t$_{2g}$ problem, while the IPT
scheme, which is very successful for  the one-band DMFT problem
\cite{review_dmft} becomes quite uncertain in the anisotropic
multi-band case \cite{ldapp}. 

\vskip .1in

{\it LDA+DMFT results and discussions -}

\begin{figure}
\includegraphics[width=7cm,keepaspectratio,angle=180]{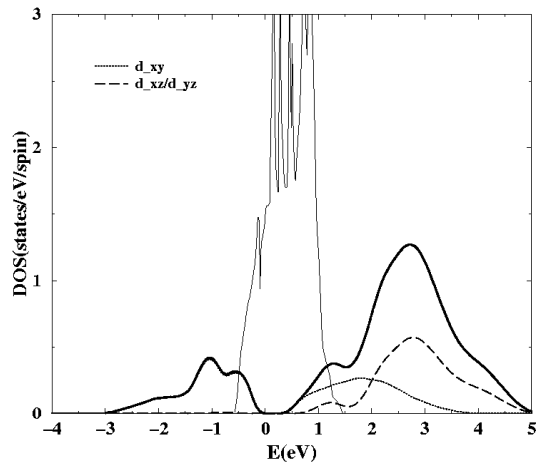}
\caption{DMFT spectral function at $T= 580 K$ (thick line) and the LDA DOS (thin line). 
$\mu$ = 0. The d$_{xy}$ and d$_{yz}$/d$_{xz}$ contributions are shown with dotted and dashed
lines respectively. The spectral weight of the broad valence states is contributed almost
entirely by d$_{xy}$.}
\label{PES}
\end{figure}

Fig.\ \ref{PES} shows the orbital-resolved and total spectral functions obtained using the above explained
LDA+DMFT technique, in comparison with the LDA total DOS. Consideration of the correlation effect 
beyond LDA within the framework of DMFT  opens up a gap of $\approx$ 0.3 eV between the occupied
and unoccupied spectra, signaling the insulating nature of the compound while the LDA DOS shows
the finite density of states at the Fermi energy. We also note the appreciable broadening of the
bandwidth in LDA+DMFT compared to the LDA bandwidth due to the redistribution of the spectral weight
following the opening of the gap. The occupied bandwidth obtained in our LDA+DMFT calculation is
in good agreement with  recent photo-emission measurements \cite{pes} which is about 2.5-3 eV.
IPT-based LDA+DMFT calculations\cite{craco} gave a bandwidth of
1.5-2 eV.
%.This is in contrast with recent IPT based DMFT calculations \cite{craco} which gives the occupied 
%band-width more like 1.5 eV.

The on-site matrix elements of the tight-binding representation of the 
 downfolded 
$t_{2g}$ Hamiltonian in the NMTO Wannier function basis show that the
$d_{xy}$ orbital energy is about 0.4 eV 
%%{\bf RV: Tanusri, Craco et
%%al. report 0.2eV which they have from the LAPW data, maybe one should
%%stress here that you are talking about the Wannier orbitals and they are
%%talking about the bare orbitals, oherwise there can be confusion} 
lower than the $d_{xz}$ and  $d_{yz}$ orbital energy, which
are degenerate within the chosen co-ordinate system. This splitting is much smaller than the total
$t_{2g}$ bandwidth of the LDA DOS which is about 2 eV and is also smaller than the individual bandwidths
of $d_{xy}$ and  $d_{yz}/d_{xz}$. As a consequence, in the LDA DOS the $d_{xy}$ orbital is occupied by only 0.49
electron with the rest occupying the $d_{xz}$ and  $d_{yz}$
orbitals. Turning to the LDA+DMFT results, the
occupation matrix shows that the $d_{xy}$ orbital becomes nearly full in
contrast to LDA. The $d_{xy}$
orbital is found to contain 0.98 electrons. The sharp increase in orbital
polarization compared to the
 LDA result is in conformity with similar LDA+DMFT studies in LaTiO$_{3}$ and YTiO$_{3}$\cite{Pavarini_04},
but is in contrast to IPT-based LDA+DMFT calculations for TiOCl which
report\cite{craco}
 non-negligible
inter-orbital mixing, with 70$\%$ of the electrons residing in the $d_{xy}$.

%===========================================================================
%===========================================================================

As mentioned already,  the gap value obtained in our calculation is about 0.3 eV, while the gap extracted
from the optical conductivity data is about 1 eV in reflectivity
experiments\cite{Caimi} 
and 2 eV in transmittance experiments\cite{Maule}, the later being claimed as
 more sensitive for the gap determination\cite{Markus}. 
 As mentioned earlier, our QMC-DMFT calculations were carried out at a temperature, T = 580 K, 
while the experimental measurements were carried out at room temperature
( $\approx$ 300K) and below . Recent 
work on V$_{2}$O$_{3}$ \cite{v2o3} shows the successive filling of the Mott-Hubbard gap on raising the 
temperature.  Therefore, the use of
temperatures higher than
  room temperature may be a cause of the underestimation 
of the gap value obtained in the theoretical result. Nevertheless, it is hard to justify such a large discrepancy
as merely a temperature effect. The IPT-based DMFT calculations which  claim\cite{craco2} to achieve
significantly lower temperatures than the QMC-DMFT calculations,
 predict
for TiOCl (with U=3.3 eV and J= 1 eV) a gap which is less
than 0.5 eV \cite{craco}. The only way to reproduce a gap value as large as $\approx$ 2 eV, within the 
single-site DMFT scheme, seems to be to use an unreasonably high U
value. We  therefore suspect that a major
contribution to this discrepancy 
is caused by the neglect of the component of the self-energy between different sites, {\it i.e.} by the 
single-site
approximation of the DMFT. The dimerization of the crystal and formation of spin-singlet bonds at low
temperature below the ordering temperature can 
%in principle induce 
have a precursor at high temperatures in form of a short-range ordering effect, 
resulting into formation of dynamical Ti-Ti singlet pairs \cite{arpes}. Consideration 
of such effects is beyond the scope of the single-site DMFT since one
needs to consider at least a cluster
of 2 Ti sites to capture these features. Experience with VO$_{2}$ \cite{Biermann} already indicates that
such effects can be extremely important in widening up the band gap. We
intend to consider the 
cluster DMFT for TiOCl in our forthcoming communication. We mention here that the phonon effects
could be important as well, as already found in connection with orbital fluctuations,\cite{Saha_04} 
leading to additional smearing  of the spectral features which is probably seen in the
experimental PES data\cite{pes}.  

{\it Acknowledgments.-}
We would like to thank R. Claessen for discussing his results with us
prior
publication as well as P. Lemmens and M. Gr\"uninger for useful
comments.
 One of us (R. V.) thanks the German Science Foundation for financial support.

%%%%%%%%%%%%%%%%%%%%%%%%%%%%%%%%%%%%%%%%%%%%%%%%%%%%%%%%%%%%%%%%%%%%%%%%%
%%%%%%%%%%%%%%%%%%%%%%%%%%%%%%%%%%%%%%%%%


\begin{thebibliography}{99}
\bibitem{arpes} The signature of such dynamical fluctuation is already evident in the angle resolved
photo-emission spectra from R. Claessen's group.
\end{thebibliography}
\end{document}